\documentclass[aps,pra,twocolumn,amsmath,amssymb,superscriptaddress]{revtex4-2}

\usepackage{graphicx}
\usepackage{textcomp} 
\usepackage{gensymb} 
\usepackage{bbold}
\usepackage{dsfont}
\usepackage{siunitx}
\usepackage[normalem]{ulem} 
\usepackage{lipsum}
\usepackage{color}
\usepackage[dvipsnames]{xcolor}

\newcommand{\euler}{\mathrm{e}}
\newcommand{\imag}{\mathrm{i}}
\newcommand{\diff}{d}
\newcommand{\expect}[1]{\left< #1 \right>}
\newcommand{\uw}{\uparrow}
\newcommand{\dw}{\downarrow}

\begin{document}

\title{Spin-Locking Spectroscopy of Harmonic Motion}

\author{Florian Kranzl}
\affiliation{Institut f\"ur Quantenoptik und Quanteninformation, \"Osterreichische Akademie der Wissenschaften, Technikerstra\ss{}e 21a, 6020 Innsbruck, Austria}
\affiliation{Institut f\"ur Experimentalphysik, Universit\"at Innsbruck, Technikerstra\ss{}e 25, 6020 Innsbruck, Austria}

\author{Adria Rospars}
\altaffiliation[Present address: ]{Ecole Normale Sup\'erieure de Lyon, CNRS, Laboratoire de Physique, F-69342 Lyon, France}
\affiliation{Institut f\"ur Quantenoptik und Quanteninformation, \"Osterreichische Akademie der Wissenschaften, Technikerstra\ss{}e 21a, 6020 Innsbruck, Austria}

\author{Johannes Franke}
\affiliation{Institut f\"ur Quantenoptik und Quanteninformation, \"Osterreichische Akademie der Wissenschaften, Technikerstra\ss{}e 21a, 6020 Innsbruck, Austria}
\affiliation{Institut f\"ur Experimentalphysik, Universit\"at Innsbruck, Technikerstra\ss{}e 25, 6020 Innsbruck, Austria}

\author{Manoj K. Joshi}
\affiliation{Institut f\"ur Quantenoptik und Quanteninformation, \"Osterreichische Akademie der Wissenschaften, Technikerstra\ss{}e 21a, 6020 Innsbruck, Austria}

\author{Rainer Blatt}

\author{Christian F. Roos}
\email{christian.roos@uibk.ac.at}
\affiliation{Institut f\"ur Quantenoptik und Quanteninformation, \"Osterreichische Akademie der Wissenschaften,
Technikerstra\ss{}e 21a, 6020 Innsbruck, Austria}
\affiliation{Institut f\"ur Experimentalphysik, Universit\"at Innsbruck, Technikerstra\ss{}e 25, 6020 Innsbruck, Austria}

\date{\today}

\begin{abstract}
    Characterization of noise of a quantum harmonic oscillator is important for many experimental platforms. We experimentally demonstrate motional spin-locking spectroscopy, a method that allows us to directly measure the motional noise spectrum of a quantum harmonic oscillator. We measure motional noise of a single trapped ion in a frequency range from $\SI{200}{Hz}$ to $\SI{5}{kHz}$ with a power spectral density that resolves noise over two orders of magnitude. Coherent modulations in the oscillation frequency of the oscillator can be probed with a relative frequency sensitivity at the $10^{-6}$~level.
\end{abstract}

\maketitle

\section{Introduction}

Controlling the motional state of trapped particles is important for a number of quantum computation and simulation platforms. Radio-frequency traps use electric fields to confine charged ions and optical tweezer arrays use focused light fields to confine neutral atoms in harmonic potentials. In any of these platforms, the particles must be cooled close to their motional ground state and their motional state must remain stable to enable quantum operations of high fidelity. In particular, two-qubit entangling operations such as the M{\o}lmer-S{\o}rensen gate~\cite{Sorensen1999, Sorensen2000} use common motional modes to mediate entanglement between two qubits. Fluctuations in the trapping potential disturb the motional state, which can lead to errors in quantum gates. It is therefore important to characterize noise that is acting on a trapped particle.

A number of techniques for measuring frequency noise of quantum harmonic oscillators describing the motion of a particle has been developed in recent years: Noise can be probed by Ramsey experiments on Fock-state superpositions~\cite{McCormick2019Nature}, motional spin-echo experiments on displaced coherent states~\cite{McCormick2019QST}, CPMG-type experiments on displaced cat states~\cite{Milne2021}, and response measurements to modulated periodic drives~\cite{Keller2021}. Furthermore, motional coherence can be measured by a Ramsey experiment on a motional sideband transition. Some of these methods give only indirect access to the noise spectrum~\cite{McCormick2019QST, McCormick2019Nature, Milne2021}, or require the experimental determination of a response function for reconstructing the spectrum~\cite{Keller2021}.

In our work, we experimentally demonstrate that spin-locking spectroscopy allows for a direct measurement of the noise spectrum of a quantum harmonic oscillator, in particular of a (harmonically) trapped ion. In a spin-locking experiment~\cite{hartmann1962nuclear}, the free-evolution period of a Ramsey experiment is replaced with a continuous drive that resonantly couples the two levels and, in the absence of noise, leaves the superposition state prepared by the first Ramsey pulse invariant. Noise with a frequency equal to the drive's Rabi frequency leads to a depolarization with a rate determined by the noise strength. Noise at other frequencies is suppressed as the drive serves as a continuous decoupling pulse for these noise terms. 
This simple response allows one to directly measure the noise spectrum by scanning the Rabi frequency over the frequency range of interest. Spin-locking spectroscopy has been used to study noise in a superconducting qubit~\cite{yan2013rotating}, detecting weak radio-frequency magnetic fields~\cite{loretz2013radio}, or measuring laser frequency noise~\cite{zhang2021estimation} or electric field noise~\cite{Bonus:2025} with a single trapped ion.

Motional spin-locking spectroscopy measures noise by detecting phase fluctuations on a transition whose energy depends on the motional frequency of the oscillator. In our experiment, the quantum harmonic oscillator is realized by a single ion trapped in a linear radio-frequency trap. It is driven by the light field of a frequency-stable laser, which continuously drives a transition involving two motional states of the trapped ion. In this work, we first give a theoretical description of spin-locking spectroscopy~(Sec.~\ref{sec:spin-locking_spectroscopy}). Then we introduce the concept of motional spin-locking spectroscopy and present the measurement of motional noise in the frequency range from $\SI{200}{Hz}$ to $\SI{5}{kHz}$. Furthermore, we detect partially coherent modulations in the oscillator frequency with a relative frequency sensitivity at the $10^{-6}$~level~(Sec.~\ref{sec:motional_spin-locking_spectroscopy}).

\section{Spin-locking spectroscopy} 
\label{sec:spin-locking_spectroscopy}

\begin{figure}
    \centering
    \includegraphics[width=86mm]{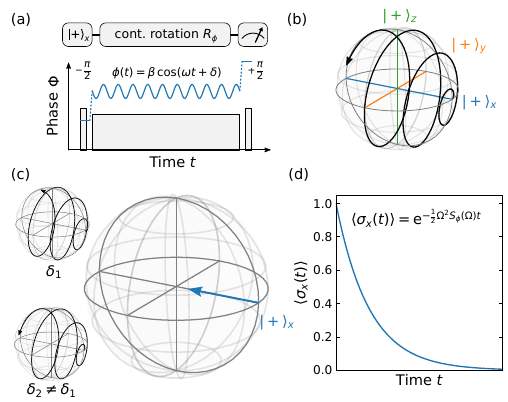}
    \caption{
    Principle of spin locking.
    (a)~In spin locking, a qubit is prepared in a superposition $\left| + \right>_x = (\left|\downarrow \right> + \left| \uparrow \right>)/\sqrt{2}$ by application of a $\pi/2$~pulse, continuously rotated around the $x$ axis and finally $\left< \sigma_x \right>$ is measured. If the phase of the rotation is modulated with a frequency that matches the Rabi frequency, then
    (b)~the state of the qubit evolves along a spiral.
    (c)~Under noise, the orientation and the inclination of the trajectory is changing with each realization of the experiment, which leads to a depolarization of the state.
    (d)~The depolarization is described by an exponential decay with time~$t$, Eq.~\eqref{eq:exponential_decay}, with a rate proportional to the phase noise power spectral density~$S_\phi$ at the Rabi frequency~$\Omega$.
    }
    \label{fig:spin-locking_principle}
\end{figure}

In spin-locking spectroscopy, a two-level system (qubit) is resonantly driven by an oscillator with Rabi frequency $\Omega$. This process is described by
\begin{equation}
    \label{eq:Hamiltonian_light_matter}
    H_\mathrm{int}(t) = \frac{\hbar \Omega}{2} \left(
        \euler^{-\imag \phi(t)} \sigma_+ +
        \euler^{\imag \phi(t)} \sigma_-
    \right).
\end{equation}
where the Hamiltonian $H_\mathrm{int}$ is given in the interaction picture with respect to the bare Hamiltonian $H_\mathrm{TLS} = \hbar \omega_0 \sigma_z / 2$ of the qubit. 
Here, the phase $\phi(t) = \int_0^t \Delta(t') dt' + \phi(0)$ accounts for temporal variations of the phase between the oscillator and the qubit that could arise from phase noise of the oscillator or transition frequency fluctuations of the qubit caused by other noise processes. This Hamiltonian can be transformed into an additional interaction picture with respect to $H_0 = (\hbar\Omega/2) \sigma_x$. For small phase variations, $\left| \phi \right| \ll 1$, and after applying the rotating-wave approximation, one obtains the Hamiltonian governing the evolution under a time-dependent phase $\phi(t)$,
\begin{equation}
    \label{eq:Hamiltonian_noise}
    H(t) = \frac{\hbar\Omega}{2} \phi(t) \left[ 
        \cos(\Omega t) \sigma_y -
        \sin(\Omega t) \sigma_z
    \right].
\end{equation}
The spin-locking protocol sketched in Fig.~\ref{fig:spin-locking_principle}(a) requires one to prepare the qubit in an eigenstate of the $\sigma_x$ operator, subjecting it to the Hamiltonian $H$ for a duration $t$ and measuring the spin projection $\sigma_x$. 
The action of the spin-locking sequence can be understood by first considering a deterministic modulation of the phase of the light field, $\phi(t) = \beta \cos(\omega t + \delta)$, where $\beta$ is the modulation index, $\omega$ the modulation frequency, and $\delta$ a phase offset. If the modulation frequency matches the Rabi frequency, $\omega = \Omega$, Eq.~(\ref{eq:Hamiltonian_noise}) shows that this frequency component will give rise to a spin rotation in this interaction picture as the multiplication with $\cos(\Omega t)$ or $\sin(\Omega t)$ results in a non-zero term after time-averaging. The state of the qubit describes a spiral on the Bloch sphere, as shown in Fig.~\ref{fig:spin-locking_principle}(b). The initial direction of the spiral is determined by the phase offset~$\delta$ and the inclination along the $x$~axis by the modulation index~$\beta$. If the modulation frequency is far off-resonant from the Rabi frequency, $\omega \neq \Omega$, the dynamics of Hamiltonian~\eqref{eq:Hamiltonian_noise} is averaged out over long times and the initial state remains unaffected. As an important consequence, a spin-locking sequence only senses signals at a frequency that coincides with the driving Rabi frequency.

If the phase of the light field is governed by noise, the evolution of the state changes. Then, $\phi(t)$ is a random variable and for each experimental realization its initial value cannot be controlled, leading to trajectories with different orientations and inclinations. When averaging over multiple realizations of the experiment, the state depolarizes along~$x$~[Fig.~\ref{fig:spin-locking_principle}(c)]. For zero-mean Gaussian noise, the magnetization $\left< \sigma_x (t) \right>$ is described by an exponential decay, with a rate proportional to the noise power spectral density at the Rabi frequency~\cite{willick2018efficient, kubo1963stochastic}, as depicted in Fig.~\ref{fig:spin-locking_principle}(d).
By varying the Rabi frequency, noise components at different frequencies can be probed.

\begin{figure*}
    \centering
    \includegraphics[width=\textwidth]{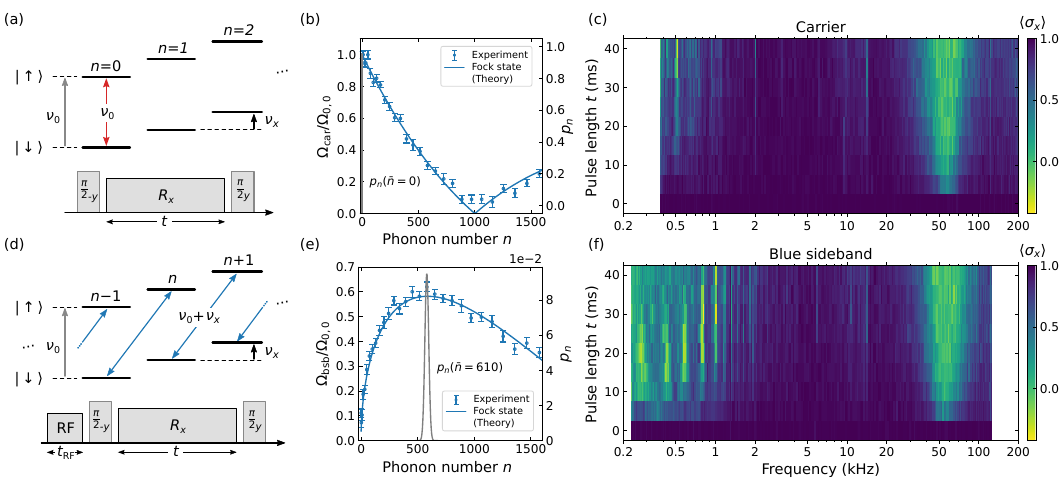}
    \caption{
    Motional spin-locking spectroscopy. 
    (a)~The carrier transition is affected only by noise on the laser or the qubit, for all Fock states~$\left| n \right>$. The spin-locking spectroscopy protocol consists of a $\pi/2$ pulse around the $-y$ axis, followed by a continuous rotation $R_x$ around the $x$ axis for a duration $t$. The expectation value $\left< \sigma_x(t) \right>$ is measured by a final analysis $\pi/2$ pulse.
    (b)~Relative coupling strength to the carrier transition as a function of the phonon number. The gray peak shows the phonon distribution for the ground-state cooled state used in the experiment. The experimental data shows the coupling strength for a coherent motional state which approximates the coupling strength of a Fock state for large phonon numbers. The Lamb-Dicke parameter is $\eta = 0.038$.
    (c)~Noise probed on the carrier transition shows a dominant noise peak at $\SI{60}{kHz}$, which originates from the laser lock feedback loop.
    (d)~Motional noise is affecting the transition frequency between two Fock states $\left| n \right>$ and $\left| n+1 \right>$. By probing the sideband transition, motional noise is detected. In the experiment, a coherent motional state is created by applying a resonant radio-frequency drive (RF) in order to maximize the coupling to the blue sindeband.
    (e)~The relative coupling strength to the blue sideband has a maximum at $\bar n = 610$~phonons. The phonon distribution of the coherent state is narrow compared to the variation of coupling strength around the maximum.
    (f)~The spin-locking signal obtained from probing the blue sideband transition shows motional noise in the low frequency region below approximately $\SI{5}{kHz}$.
    }
    \label{fig:motional_noise}
\end{figure*}

The evolution under a noisy Hamiltonian is obtained by the theory of stochastic Liouville equations~\cite{kubo1963stochastic}. The expectation value of the density matrix after an evolution time $t$ is $\left< \hat \rho(t) \right> = \Phi(t) \hat \rho(0)$, where $\hat \rho = (\rho_{11}, \rho_{12}, \rho_{21}, \rho_{22})^\mathrm{T}$ is the vectorized density matrix and $\Phi(t)$ is the decay operator. The decay operator contains the information on the noise power spectral density and the filter function of the spin-locking sequence (details see Appendix~\ref{sec:appendix_evolution_under_noise}). Importantly, the filter function exhibits a narrow peak that probes only noise at the driving Rabi frequency $\Omega$, while suppressing all other noise components.

The outcome of the experiment depends on the type of noise. In general, a random variable can consist of a stochastic and a deterministic part, $\phi = \phi_\mathrm{s} + \phi_\mathrm{d}$. For stationary, zero-mean Gaussian noise $\phi = \phi_\mathrm{s}$, the magnetization $\left< \sigma_x(t) \right>$ decreases exponentially with time~$t$~\cite{willick2018efficient, kubo1963stochastic},
\begin{equation}
    \label{eq:exponential_decay}
    \left< \sigma_x(t) \right> = \euler^{- \frac{1}{2} \Omega^2 S_\phi(\Omega) t} \left< \sigma_x(0) \right>,
\end{equation}
where the rate of decay is determined by the phase noise power spectral density $S_\phi(\Omega)$. 
The noise power spectral density is related to the autocorrelation function $C_\phi(t_1-t_2) = \left< \phi(t_1) \phi(t_2) \right>$ of the underlying random variable by the Wiener-Khinchin theorem, 
$S_\phi(\omega) = \int_{-\infty}^{\infty} C_\phi(t) \euler^{-\imag \omega t} \diff t$. The frequency noise power spectral density is linked to the phase noise power spectral density by $S_\nu(\omega) = \omega^2 S_\phi(\omega)$~\cite{cutler1966frequency}.

In the case where the phase is varying deterministically, $\phi = \phi_\mathrm{d}$, the phase may be described by a coherent modulation $\phi(t) = \sum_k \beta_k \cos \left( \omega_k t + \delta_k \right)$, where each modulation frequency $\omega_k$, $k=1,\dots,N$, is weighted by a modulation index $\beta_k$. A spin-locking experiment that is carried out for such a phase modulation leads to coherent oscillations in the magnetization. Consider a qubit that is initially prepared in~$\left| + \right>_x$. If the drive frequency is resonant with one of the modulation frequencies, $\Omega = \omega_k$, and is far-detuned from all other modulation frequencies, $\left| \Omega - \omega_{k' \neq k} \right| \gg \beta_{k'} \Omega$, the magnetization along~$x$ is~(Appendix~\ref{sec:appendix_coherent_modulation})
\begin{equation}
    \left< \sigma_x(t) \right> = \cos \left( \frac{1}{2} \beta \Omega t \right).
\end{equation}

In the case where both random noise and a coherent modulation change the phase $\phi = \phi_\mathrm{s} + \phi_\mathrm{d}$, the magnetization along~$x$ is described by damped oscillations (Appendix~\ref{sec:appendix_coherent_modulation_with_noise}),
\begin{equation}
    \label{eq:damped_oscillations}
    \left< \sigma_x(t) \right> = \cos\left( \frac{1}{2} \beta \Omega t \right) \euler^{ -\frac{1}{2} \Omega^2 S_\phi(\Omega) t }.
\end{equation}

In our experiments, we implement the spin-locking protocol with $^{40}$Ca$^+$ ions held in a linear radiofrequency trap, with transverse trapping frequencies of approximately $\SI{3}{MHz}$.  The qubit is encoded in a linear combination of the Zeeman ground state $\left| \dw \right> = \left| 4^2\mathrm{S}_{1/2}, m=1/2 \right>$ and the metastable Zeeman state $\left| \uw \right> = \left| 3^2\mathrm{D}_{5/2}, m=5/2 \right>$. The quadrupole transition connecting these two states is driven by a Ti:Sa laser at $\SI{729}{nm}$ stabilized to a high-finesse cavity, resulting in a linewidth of less than $\SI{10}{Hz}$. The $k$ vector of the light field encloses an angle of approximately $\SI{45}{\degree}$ with both transverse axes $x$ and $y$, leading to an overlap with both motional modes. The degeneracy of the trapping frequencies is lifted by applying a constant electric field, which leads to trapping frequencies of $\nu_x = 2\pi\times\SI{3.167}{\MHz}$ along the $x$ axis and $\nu_y = 2\pi\times\SI{2.909}{\MHz}$ along the $y$ axis. This allows us to probe noise of the two transverse motional modes individually. Further details about the experimental apparatus can be found in Ref.~[\onlinecite{Kranzl:2022}].

Fig.~\ref{fig:motional_noise}(a) shows the experimental protocol used for spin-locking spectroscopy. First, a $\pi/2$ pulse creates a superposition $\left| +\right>_x = (\left| \downarrow \right> + \left|\uparrow \right>) / \sqrt{2}$. Then, a spin-locking pulse of duration~$t$ and Rabi frequency~$\Omega$ probes noise that affects the carrier transition. Finally, the remaining magnetization $\left< \sigma_x(t) \right>$ is measured by an analysis $\pi/2$ pulse around the $y$-axis followed by a fluorescence measurement in order to rotate the $\left| \pm \right>_x$ eigenstates into the fluorescence measurement basis.  A decay of $\left< \sigma_x(t) \right>$ indicates the presence of noise with frequency $\Omega$. The carrier transition is affected by noise on the laser, as well as by noise on the qubit transition frequency~$\nu_0$ itself.

Since the coupling strength depends on the phonon number, driving the spin-locking pulse on a hot ion would lead to a dispersion of spectral features. Therefore, the measurements were carried out for a ground-state cooled ion ($\bar n \approx 0.1$), which maximizes the coupling strength for the carrier transition~[Fig.~\ref{fig:motional_noise}(b)]. For the measurement of the coupling strength the mean phonon number~$\bar n$ was controlled by bringing the ion into a coherent motional state with a resonant radio-frequency drive, as described in detail in Sec.~\ref{sec:motional_spin-locking_spectroscopy}.

The experimentally measured spin-locking signal of the carrier transition is shown in Fig.~\ref{fig:motional_noise}(c). At $\SI{60}{kHz}$, there is a dominant noise peak which originates from the feedback loop of the $\SI{729}{nm}$ laser lock. Additionally, there are a number of narrow noise peaks at lower frequencies. The peak at $\SI{2}{kHz}$, for instance, is also created by the laser lock, namely by a modulation of an etalon inside the laser cavity.

Similar experiments were previously reported in Ref.~\cite{zhang2021estimation}. Here, we extend the method to the detection of frequency noise in the motion of the trapped ion.

\section{Motional spin-locking spectroscopy}
\label{sec:motional_spin-locking_spectroscopy}

\begin{figure*}
    \centering
    \includegraphics[width=\textwidth]{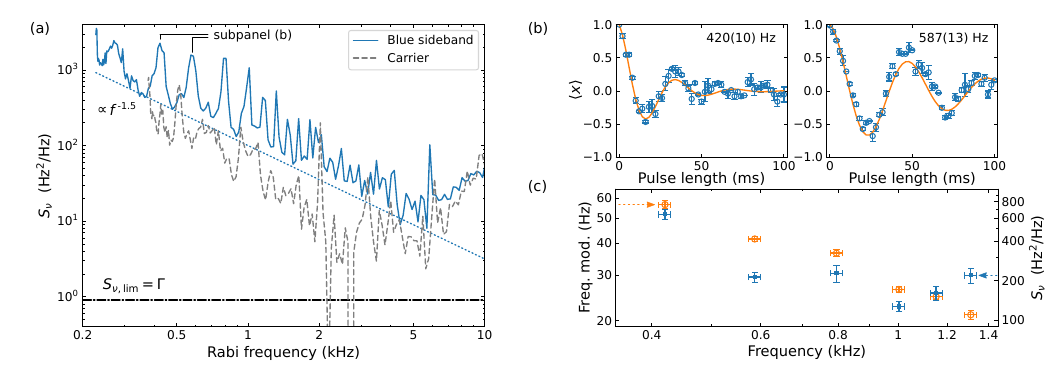}
    \caption{
    Frequency noise spectrum of a quantum harmonic oscillator.
    (a)~The frequency noise power spectral density $S_\nu$ of the blue sideband transition and of the carrier transition are shown. The dotted line is a power-law function as a guide to the eye. Spin-locking spectroscopy is limited by the spontaneous decay rate $\Gamma = \SI{0.9}{\per\s}$ of the $3d{}^2D_{5/2}$ level to noise above $S_{\nu,\mathrm{lim}} = \Gamma$, as indicated by the dash-dotted line.
    (b)~The peaks highlighted in subpanel (a) show a considerable degree of coherence, which leads to damped oscillations of the magnetization $\left< \sigma_x(t) \right>$. The solid lines are fits to Eq.~\eqref{eq:damped_oscillations}, which gives
    the frequency modulation (orange) and the noise spectral density (blue) for a series of peaks that are shown in subpanel (c).
    }
    \label{fig:noise_spectrum}
\end{figure*}

The motional noise spectroscopy protocol is based on continuously driving a transition connecting two different eigenstates of the quantum-harmonic oscillator describing the motion of a trapped ion. On the sideband transition depicted in Fig.~\ref{fig:motional_noise}(d), the light field couples states $\left| \dw, n \right>$ with states $\left| \uw, n+1 \right>$. 
Therefore, frequency noise of the oscillator will give rise to noise on the transition frequency on top of the noise already present on the carrier transition.  
However, the transition matrix elements strongly depend on the motional state the ion is prepared in. To avoid a dispersion of coupling strength, which would adversely affect our capability of sensing noise at the frequency set by the Rabi frequency, one could prepare the ion in its motional ground state by sideband laser cooling. Doing so would, however, reduce the Rabi frequency by the Lamb-Dicke parameter $\eta$ as compared to the carrier Rabi frequency, which would strongly reduce our capability of sensing noise at higher frequencies. 

Instead of trying to prepare higher Fock states of motion, here, we overcome this problem by following a much simpler approach: using an electric field oscillating at the ion's oscillation frequency, we prepare a coherent state of motion (outside the Lamb-Dicke regime) with a mean phonon number that maximizes the sideband coupling strength. Although this coherent state is represented by a superposition of Fock states, the variance of the transition matrix elements of those states with substantial population is tiny for $\eta\ll 1$. Note that this condition does not constitute a serious restriction as $\eta$ can be made arbitrarily small by making the wave vector of the laser beam nearly perpendicular to the direction of ion motion or by replacing a single ion with an $N$-ion crystal for which $\eta\propto 1/\sqrt{N}$.

In order to optimally displace the motional state, we first determine the average sideband Rabi frequency $\bar{\Omega}_{BSB}(\bar{n})=\sum_n p_n(\bar{n})\Omega_{n,n+1}$ as a function of the motional state's displacement. Here, $\bar{n}$ denotes the average phonon number and $p_n(\bar{n})$ the occupation probability of the Fock state $|n\rangle$ for a coherent state with $\bar{n}$ phonons. For this measurement, an initially ground-state cooled ion in the state $\left| \downarrow \right>$ is brought into a coherent motional state by applying a radio-frequency drive that is resonant with the transverse mode frequency~$\nu_x$ for a duration~$t_\mathrm{RF}$. Then, a light pulse of fixed length~$t_p$ is resonantly applied on the blue sideband transition. The probability of finding the ion in the excited state is approximately described by $p_\uparrow = \sin^2 \left( \bar{\Omega}_{BSB}(\bar{n}) t_p/2 \right)$. This approximation assumes that the phonon distribution~$p_n(n)$ is narrow compared to the rate of change in coupling strength. The average sideband Rabi frequency~$\bar{\Omega}_{BSB}(\bar{n})$ is then extracted from the measured excitation probability~$p_\uparrow$.

Fig.~\ref{fig:motional_noise}(e) shows 
the experimentally measured average sideband Rabi frequency normalized to the carrier Rabi frequency $\Omega_{0,0}$ of a ground-state cooled ion. The measurements are in good agreement with the theoretically calculated normalized Rabi frequency for $\eta=0.038(3)$. At the peak of the curve, we expect the variance of Rabi frequencies, $\mbox{Var}(\Omega_{BSB})=\overline{\Omega^2}_{BSB}-(\overline{\Omega}_{BSB})^2$, to be very small. In our case the spread of relative Rabi frequencies evaluates to only $\sigma_\Omega=\sqrt{\mbox{Var}(\Omega_{BSB})}/\Omega_{0,0}=3 \times 10^{-4}$. This narrow distribution enables us to neglect the phonon-number dependence and to describe the light-ion interaction simply by Hamiltonian~\eqref{eq:Hamiltonian_light_matter} and~\eqref{eq:Hamiltonian_noise} of a resonantly driven two-level atom in the interaction picture.

The experimental protocol for motional spin-locking spectroscopy is shown in Fig.~\ref{fig:motional_noise}(d). First, a ground-state cooled ion is prepared in a coherent motional state, $\left| \alpha \right> = \sum_0^\infty c_n \left| \dw, n \right>$ with $c_n = (\alpha^n / \sqrt{n!}) \exp(-|\alpha|^2/2)$, by applying a resonant radio-frequency drive. The mean phonon number $\bar n = |\alpha|^2$ is set to maximize the coupling to the blue sideband. A $\pi/2$ pulse on the blue sideband creates a superposition $\left| + \right>_x = \sum_0^\infty c_n ( \left| \dw , n \right> + \left| \uw, n + 1 \right> ) / \sqrt{2}$. Then, a continuous rotation $R_x$ around the $x$ axis with Rabi frequency $\Omega$ is driven on the blue sideband for a duration $t$. During this driving period under Hamiltonian~\eqref{eq:Hamiltonian_noise}, variations in the phase $\phi(t)$ are sensed by the qubit. Finally, a $\pi/2$ pulse measures the remaining magnetization $\left< \sigma_x(t) \right>$. 

There are several noise processes present in the experiment: Laser noise and magnetic field noise affect both the electronic and the motional transition, while trapping noise only affects the motional transition. To distinguish between the different noise processes, the noise spectroscopy is carried out on two different transitions: the motional sideband transition and the carrier transition. This allows us to identify noise components that are only present on the motional state, and do not arise from laser frequency noise or magnetic field noise. Figures~\ref{fig:motional_noise}(c) and (f) compare noise on the carrier transition with noise on the motional sideband transition. The motional noise spectrum shows a series of distinct peaks that are only present on the motional sideband transition and are therefore caused by fluctuations in the trapping frequency. 

The measured motional noise spectrum is presented in Fig.~\ref{fig:noise_spectrum}(a). The frequency noise spectral density $S_\nu$ of the trapping frequency is obtained from a least-squares fit of Eq.~\eqref{eq:exponential_decay} to the measured temporal evolution of the magnetization. In this first step, we neglect coherent properties of the frequency fluctuations to give an estimate of $S_\nu$. The motional frequency noise spectral density is dominated by a series of peaks in the range from $\SI{200}{\Hz}$ to a few kHz. In this region, the noise background of $S_\nu(\omega)$ drops approximately as $\omega^{-1.5}$. Above $\SI{10}{\kHz}$, laser noise becomes dominant. In particular, the noise peak in the carrier spectrum around $\SI{60}{\kHz}$ is caused by the servo bump of the frequency lock of our laser system [Fig.~\ref{fig:motional_noise}(e)].

The motional noise peaks show considerable coherence. In Figure~\ref{fig:noise_spectrum}(b), the time evolution of the magnetization~$\left< \sigma_x(t) \right>$ is shown for a few of the peaks. For these peaks, detailed time-evolution data was measured. Instead of a simple exponential decay, the magnetization exhibits damped oscillations. The modulation index and the noise spectral density is then obtained for individual frequencies in a range between approx. 400 and $\SI{1400}{\Hz}$ from fits of Eq.~\eqref{eq:damped_oscillations} to the measured time evolution. The results are presented in Fig.~\ref{fig:noise_spectrum}(c). In the evaluated range, the trapping frequency is modulated by up to $\Delta \nu = \SI{56(2)}{\Hz}$, corresponding to relative changes of $\Delta \nu / \nu_x = \SI{2e-5}{}$ with an uncertainty in the order of~$10^{-6}$.
A potential source of the coherent modulation of the trapping frequency could be the electronic stabilization circuit of the trap drive signal.

The accessible frequency range of approx. $\SI{200}{Hz}$ to $\SI{5}{kHz}$ is limited by two main factors: For low frequencies, the present noise becomes strong and is outside the weak-noise limit~\cite{willick2018efficient}. For high frequencies, laser noise becomes dominant and superimposes potential motional noise. Throughout this frequency range, the measured noise spectrum does not yet reach the limit $S_{\nu,\mathrm{lim}}$ set by the finite lifetime of the metastable state.

\section{Conclusion}
\label{sec:conclusion}

In this work, we have introduced a motional spin-locking spectroscopy method. We used this method to measure the motional noise spectrum in a trapped-ion system by continuously driving a superposition of two motional states. 

The sensitivity of spin-locking spectroscopy is limited by noise of the laser that is used to probe the motional transition. This problem could be circumvented by driving a Raman transition, which connects the two ground states of ${}^{40}\mathrm{Ca}^+$ via the metastable state so that laser noise that is common to the two paths of the Raman transition cancels. 

An additional limitation arises from the finite lifetime of the excited state $\left| \uparrow \right>$. A decay rate of $\Gamma$ leads to a lower limit of the detectable noise spectral density of $S_{\nu, \mathrm{lim}} = \Gamma$. In the case of ${}^{40}\mathrm{Ca}$, the excited state $\left| \uparrow \right> = \left| 3{}^2\mathrm{D}_{5/2}, m=5/2 \right>$ decays with a rate of $\Gamma = \SI{0.9}{\per\s}$~\cite{Barton2000lifetime}. Again, using a ground-state qubit that is connected by a Raman transition could offer a solution to eliminate a finite-lifetime limitation.

Our work provides a tool for improving the trapping stability in charged-particle traps. It could be applied to other systems with external degrees of freedom, such as optical traps.

\begin{acknowledgements}
The research work presented here has received funding under Horizon Europe programme HORIZON-CL4-2022-QUANTUM-02-SGA via the project 101113690 (PASQuanS2.1) and Institut f\"ur Quanteninformation GmbH.
\end{acknowledgements}

\appendix

\section{Evolution under noise}
\label{sec:appendix_evolution_under_noise}

The evolution of a resonantly driven qubit under a noisy phase $\phi(t)$ is described in the interaction picture by Hamiltonian~\eqref{eq:Hamiltonian_noise},
\begin{equation}
    \label{eq:Hamiltonian_noise_appendix}
    H(t) = \frac{\hbar\Omega}{2} \phi(t) \left[ 
        \cos(\Omega t) \sigma_y -
        \sin(\Omega t) \sigma_z
    \right].
\end{equation}
The evolution of the state's density matrix $\rho$ is given by the von Neumann equation, $\frac{\diff}{\diff t} \rho(t) = \frac{1}{\imag \hbar} [H(t), \rho(t)]$. In order to use the theory of stochastic Liouville equations~\cite{kubo1963stochastic}, the action of the commutator is expressed as a superoperator, giving
\begin{equation}
    \label{eq:Liouville_eq}
    \frac{\diff}{\diff t} \hat \rho(t) = \operatorname{\mathcal{L}} \hat \rho(t),
\end{equation}
where $\mathcal{L} = \frac{1}{\imag \hbar} (H \otimes \mathds{1} - \mathds{1} \otimes H^\mathrm{T})$ is the Liouville operator and $\hat \rho = (\rho_{1,1},\rho_{1,2},\rho_{2,1},\rho_{2,2})^\mathrm{T}$ denotes the vectorized density matrix. 

The solution to Eq.~\eqref{eq:Liouville_eq} can be expressed with the time-ordered exponential,
\begin{equation}
    \hat \rho(t) = \mathcal{T} \exp{ \left( \int_0^t \mathcal{L}(t') \, \diff t' \right) } \hat \rho(0) .
\end{equation}
The expectation value of the time-evolved state is
\begin{equation}
    \label{eq:Phi_toExp_Liouvillian}
    \expect{\hat \rho(t)} 
    = \expect{ \mathcal{T} \exp{ \left( \int_0^t \mathcal{L}(t') \, \diff t' \right) } } \hat \rho(0)
    \equiv \Phi(t) \hat \rho(0),
\end{equation}
with the relaxation operator $\Phi(t)$. The relaxation operator $\Phi(t)$ can be expressed by the cumulant expansion,
\begin{align}
    \label{eq:Phi_toExp}
    \Phi(t) = \mathcal{T} & \exp \Bigg(
    \int_0^t \diff t_1 \expect{\mathcal{L}(t_1)}_\mathrm{c} \nonumber \\
    & + \int_0^t \diff t_1 \int_0^{t_1} \diff t_2 \expect{\mathcal{L}(t_1) \mathcal{L}(t_2)}_\mathrm{c} 
    + \dots
    \Bigg),
\end{align}
where $\expect{\cdot}_\mathrm{c}$ denote the cumulants. The cumulants can be determined by comparing the series expansions of Eq.~\eqref{eq:Phi_toExp_Liouvillian} and~\eqref{eq:Phi_toExp}, yielding $\expect{\mathcal{L}(t_1)}_\mathrm{c} = \expect{\mathcal{L}(t_1)}$ and $\expect{\mathcal{L}(t_1) \mathcal{L}(t_2)}_\mathrm{c} = \expect{\mathcal{L}(t_1) \mathcal{L}(t_2)} - \expect{\mathcal{L}(t_1)} \expect{\mathcal{L}(t_2)}$ for the lowest-order terms~\cite{chaturvedi1979projection}.

For a zero-mean random process, the first cumulant vanishes and the time-ordered exponential is approximated by the exponential. When considering only up to the second cumulant, the decay operator is
\begin{equation}
    \label{eq:Phi_exp}
    \Phi(t) \approx \exp \left( \int_0^t \diff t_1 \int_0^{t_1} \diff t_2 \expect{\mathcal{L}(t_1) \mathcal{L}(t_2)} \right).
\end{equation}

For Hamiltonian~\eqref{eq:Hamiltonian_noise_appendix}, the Liouville operator evaluates to
\begin{equation}
    \mathcal{L}(t) = \frac{1}{\imag} \frac{\Omega}{2} \phi(t)
    \begin{pmatrix}
        0 & -\imag c(t) & -\imag c(t) & 0 \\
        \imag c(t) & -2 s(t) & 0 & -\imag c(t) \\
        \imag c(t) & 0 & 2 s(t) & -\imag c(t) \\
        0 & \imag c(t) & \imag c(t) & 0
    \end{pmatrix}
\end{equation}
with the abbreviations $c(t) \equiv \cos(\Omega t)$ and $s(t) \equiv \sin(\Omega t)$. The integral in Eq.~\eqref{eq:Phi_exp} can be evaluated by making use of the Wiener-Khinchin theorem, $\expect{\phi(t_1) \phi(t_2)} = \frac{1}{2\pi} \int_{-\infty}^{+\infty} \diff \omega \, \euler^{\imag \omega (t_2 - t_1)} S_\phi(\omega)$, yielding the decay operator in the long-time limit, $t \rightarrow \infty$,
\begin{equation}
    \label{eq:decay_operator_approximation}
    \Phi(t) = \exp \left[
        -\chi_2(t)
        \begin{pmatrix}
            1 & 0 & 0 & -1 \\
            0 & 3 & 1 & 0 \\
            0 & 1 & 3 & 0 \\
            -1 & 0 & 0 & 1
        \end{pmatrix}
    \right],
\end{equation}
where $\chi_2(t) = \frac{\Omega^2}{4} S_\phi(\Omega) \frac{t}{2}$. The expectation value of the density matrix is then
\begin{widetext}
\begin{align}
    &\expect{\hat \rho(t)} = \Phi(t) \hat \rho(0) 
    = \frac{1}{2} \begin{pmatrix}
        1+\euler^{-2\chi_2(t)} & 0 & 0 & 1-\euler^{-2\chi_2(t)} \\
        0 & \euler^{-2\chi_2(t)}+\euler^{-4\chi_2(t)} & -\euler^{-2\chi_2(t)}+\euler^{-4\chi_2(t)} & 0 \\
        0 & -\euler^{-2\chi_2(t)}+\euler^{-4\chi_2(t)} & \euler^{-2\chi_2(t)}+\euler^{-4\chi_2(t)} & 0 \\
        1-\euler^{-2\chi_2(t)} & 0 & 0 & 1+\euler^{-2\chi_2(t)} \\
    \end{pmatrix} \hat \rho(0).
\end{align}
\end{widetext}
From this result we can calculate the expectation value of the magnetization
\begin{align}
    \label{eq:mag_X_exp_decay}
    \expect{\sigma_x(t)} &= \operatorname{Tr} \left[ \expect{\rho(t)} \sigma_x  \right] = \expect{\rho_{1,2}(t)} + \expect{\rho_{2,1}(t)} \\
    &= \euler^{-4 \chi_2(t)} \left[ \rho_{1,2}(0) + \rho_{2,1}(0) \right] \\
    &= \euler^{-\frac{1}{2} \Omega^2 S_\phi(\Omega) t} \expect{\sigma_x(0)}.
\end{align}

\section{Coherent modulation}
\label{sec:appendix_coherent_modulation}

We consider a coherent phase modulation of the light driving the spin-locking sequence. The phase between light and qubit is modulated at a modulation frequency $\omega$ and a modulation index $\beta$,
\begin{equation}
    \phi(t) = \beta \cos(\omega t + \delta),
\end{equation}
where $\delta$ is the modulation phase offset. In the following, we assume the modulation index to be small, $\left| \beta \right| \ll 1$. Then, the evolution of the qubit is described in the interaction picture by Hamiltonian~\eqref{eq:Hamiltonian_noise} which takes the form
\begin{align}
    H &= \hbar \Omega^{(1)} \cos(\omega t + \delta) \left[ 
        \cos(\Omega t) \sigma_y -
        \sin(\Omega t) \sigma_z
    \right] \\
    &\approx \hbar \frac{\Omega^{(1)}}{2}  
    \left\{ 
        \cos[(\omega - \Omega) t + \delta] \sigma_y + \right. \nonumber \\
        & \hspace{30mm} \left. \sin[(\omega - \Omega) t + \delta] \sigma_z
    \right\},
\end{align}
where $\Omega^{(1)} = \frac{1}{2} \beta \Omega$ is the Rabi frequency of the 1${}^\mathrm{st}$ order modulation sideband. In the second step, we have applied a rotating-wave approximation, neglecting terms oscillating at $\Omega + \omega$.

When the Rabi frequency is made resonant with the modulation frequency, $\Omega=\omega$,  this leads to the Hamiltonian of a coherent, small phase modulation at resonance,
\begin{equation}
    H = \hbar \frac{\Omega^{(1)}}{{2}} 
        [ \cos(\delta) \sigma_y + \sin(\delta) \sigma_z ].
\end{equation}
This time-independent Hamiltonian has the time-evolution operator
\begin{multline}
    \label{eq:evol_op_coherent_mod}
    U(t) = \cos\left( \frac{\Omega^{(1)}t}{2} \right) \mathds{1} \\
    - \imag \sin\left( \frac{\Omega^{(1)}t}{2} \right) \left[
        \cos(\delta) \sigma_y + \sin(\delta) \sigma_z
    \right]
\end{multline}
and leads to the evolution of magnetization
\begin{align}
    \label{eq:mag_X_coherent}
    &\left< \sigma_x(t) \right> = \cos \big( \Omega^{(1)} t \big), \\
    \label{eq:mag_Y_coherent}
    &\left< \sigma_y(t) \right> = \sin \big( \Omega^{(1)} t \big) \sin \big( \Omega t + \delta \big), \\
    \label{eq:mag_Z_coherent}
    &\left< \sigma_z(t) \right> = -\sin \big( \Omega^{(1)} t \big) \cos \big( \Omega t + \delta \big).
\end{align}

\begin{figure}
    \centering
    \includegraphics[width=86mm]{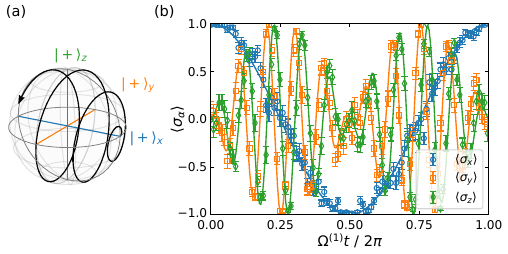}
    \caption{
    Spin-locking spectroscopy under coherent phase modulation. A superposition $(\left| 1 \right> + \left| 0 \right>) / \sqrt{2}$ on the carrier transition is driven while the light is phase modulated at $\Omega = 2\pi \times \SI{5}{\kHz}$ with a modulation index of $\beta = \SI{200}{mrad}$. The Rabi frequency of the 1$^\mathrm{st}$~order modulation sideband is $\Omega^{(1)} \approx \frac{1}{2} \beta \Omega$. 
    (a)~The evolution of the state is described by a spiral on the Bloch sphere.
    (b)~Magnetization~$\left< \sigma_\alpha \right>$ ($\alpha = x,y,z$) as a function of the spin-locking time. Points represent experimental data averaged over 150 repetitions, solid lines are theory curves using Eqs.~\eqref{eq:mag_X_coherent}-\eqref{eq:mag_Z_coherent}. The modulation phase offset $\delta$ is obtained from a least-squares fit to the data.
    }
    \label{fig:coherent_modulation}
\end{figure}

An example of the evolution under coherent phase modulation is shown in Fig.~\ref{fig:coherent_modulation}. We carried out a spin-locking experiment on the carrier transition of a ground-state cooled ion. The light that was driving the spin-locking sequence was phase-modulated at $\omega=2\pi\times\SI{5}{\kHz}$ at a modulation index of $\beta = \SI{200}{mrad}$. This particular modulation frequency
was chosen as our laser has 
the lowest frequency noise around this frequency. The observed evolution of magnetization is described well by equations~\eqref{eq:mag_X_coherent}-\eqref{eq:mag_Z_coherent}. The magnetization $\left< \sigma_x(t) \right>$ oscillates at a frequency $\Omega^{(1)}$. The magnetizations $\left< \sigma_y(t) \right>$ and $\left< \sigma_z(t) \right>$ oscillate at a frequency~$\Omega$, with an amplitude envelope of frequency~$\Omega^{(1)}$.
In the experiment, the phase relation $\delta$ between modulation signal and light is kept constant but its value is unknown. We obtained $\delta = \SI{1.48(2)}{rad}$ from a least-squares fit to the experimental data.

\section{Coherent modulation with noise}
\label{sec:appendix_coherent_modulation_with_noise}

In the case where both a coherent modulation and noise are present, the Hamiltonian is written as a sum of a deterministic part $H_\mathrm{d}$ and a stochastic part $H_\mathrm{s}$,
\begin{align}
    H(t) 
    &= H_\mathrm{d} + H_\mathrm{s}(t) \\
    &= \hbar \frac{\Omega^{(1)}}{{2}} 
        [ \cos(\delta) \sigma_y + \sin(\delta) \sigma_z ] \nonumber \\
    & \quad + \hbar \frac{\Omega}{2} \phi(t) \left[
        \cos(\Omega t) \sigma_y -
        \sin(\Omega t) \sigma_z
    \right].
\end{align}
The stochastic part is transformed into the interaction picture
\begin{equation}
    H_{\mathrm{s},\mathrm{I}}(t) = U_0(t)^\dag H_\mathrm{s}(t) U_0(t),
\end{equation}
with $U_0(t)$ given by Eq.~\eqref{eq:evol_op_coherent_mod}. This operator is of the form $H_{1,\mathrm{I}}(t) = \phi(t) (a(t) \sigma_x + b(t) \sigma_y + c(t) \sigma_z)$, where $\phi(t)$ is again assumed to be a zero-mean random process. Therefore, the evolution of the density matrix in the interaction picture $\rho_\mathrm{I}(t)$ is described by the stochastic Liouville equation in the same way as described in Sec.~\ref{sec:appendix_evolution_under_noise}. The cumulant $\expect{\mathcal{L}(t_1) \mathcal{L}(t_2)}_\mathrm{c}$ contains a large number of terms; however, in the limit $\Omega^{(1)} \rightarrow 0$ we find that $\Phi(t)$ is equal to~\eqref{eq:decay_operator_approximation}. If the qubit is initially prepared in the $\left| + \right>_x$ state, the magnetization evolves as
\begin{align}
    \expect{\sigma_x(t)} 
    &= \operatorname{Tr} \left[
        U_0(t) \expect{\rho_\mathrm{I}(t)} U_0^\dag(t) \sigma_x    
    \right] \\
    &= \operatorname{Tr} \left[
        U_0(t) \frac{1}{2} (\mathds{1} + \euler^{-4\chi_2(t)}\sigma_x) U_0^\dag(t) \sigma_x    
    \right] \\
    &= \frac{1}{2} \euler^{-4\chi_2(t)} \operatorname{Tr} \left[
        U_0(t) \sigma_x U_0^\dag(t) \sigma_x
    \right].
\end{align}
Since we have assumed $\rho_0 = \frac{1}{2} (\mathds{1} + \sigma_x)$, we can express $\sigma_x$ by $\rho_0$ and get
\begin{align}
    \label{eq:magnetization_coherent_and_noise}
    \expect{\sigma_x(t)} 
    &= \frac{1}{2} \euler^{-4\chi_2(t)} \operatorname{Tr} \left[
        U_0(t) (2\rho_0 - \mathds{1}) U_0^\dag(t) \sigma_x
    \right] \\
    &= \euler^{-4\chi_2(t)} \expect{\sigma_x(t)}_0 \\
    &= \euler^{-\frac{1}{2} \Omega^2 S_\phi(\Omega) t} \cos(\Omega^{(1)}t).
\end{align}
In the second-to-last step, $\expect{\sigma_x(t)}_0 \equiv \operatorname{Tr} [U_0(t) \rho_0 U_0^\dag(t) \sigma_x ]$ denotes the magnetization under coherent evolution, as described by Eq.~\eqref{eq:mag_X_coherent}. The main result of this calculation is that a coherent drive with noise is simply described by the product of the incoherent part and the coherent part, leading to damped oscillations.

\bibliography{main} 

\end{document}